\documentclass[preprint,amsmath,amssymb,aps]{revtex4-2}
\usepackage{graphicx}% Include figure files
\usepackage{dcolumn}% Align table columns on decimal point
\usepackage{bm}% bold math
\usepackage{hyperref}% add hypertext capabilities
\usepackage{natbib}
\usepackage{comment}
\usepackage{xcolor}

\begin{document}

\preprint{}

\title{Internal geometries regulate the symmetry of defect configurations in cell populations confined to domains with a negative Euler characteristic}% Force line breaks with \\

\author{Mina Kamao}
\author{Hiroyuki Miyoshi}
\author{Takaaki Nara}
\author{Hiroki Miyazako}
\email{hiroki\_miyazako@ipc.i.u-tokyo.ac.jp}
\affiliation{%
Graduate School of Information Science and Technology\\
The University of Tokyo
}%

\date{\today}% It is always \today, today,

\begin{abstract}

Nematic order of confined cell populations plays an important role in determining cell alignment and stable configurations of topological defects, which are related to various biomechanical phenomena. 
Topological charges (or winding numbers) of topological defects strictly depend on the Euler characteristic of the confining domain, which has typically been non-negative in studies focused on domains without internal obstacles.
However, biological tissues often surround two or more internal obstacles or holes, which inherently generate defects with negative charges.
To understand the mechanical interaction between cellular tissue and obstacles, it is necessary to elucidate the geometrical effects of obstacles on cell alignment and defects with negative charges.
Here, we investigate how cell populations achieve stable defect configurations of two $-1/2$ defects in a triply connected domain.
First, we present experimental observations of C2C12 myoblasts confined by two circular obstacles of varying diameter, demonstrating that two $-1/2$ defects are the most frequent configuration when the obstacles are sufficiently large.
Second, to theoretically validate these experimental observations, we perform systematic stability analyses of defect configurations using an explicit expression of cell alignment and numerical minimization of the Frank elastic energy.
Our numerical calculations reveal that the most stable configuration shifts continuously from a horizontal, through off-axis, to a vertical configuration as the obstacle size increases. In addition, the experimentally observed defect positions agreed with these theoretical predictions to within 60 \textmu m.
These findings suggest that obstacle sizes control the symmetry of cell alignment, providing insights into how geometric and topological constraints can generate complex force patterns during morphogenesis or organ movements.

%\begin{description}
%\end{description}
\end{abstract}

\maketitle

\section{Introduction}

Active nematics are systems of active matter in which elongated constituent units behave like self-propelled nematic liquid crystals through energy consumption and exhibit collective behavior.
Recent studies have demonstrated that active nematics are promising for physical models of biophysical phenomena \cite{kemkemer_2000_ElasticPropertiesNematoid, doostmohammadi_2018_ActiveNematics, balasubramaniam_2022_ActiveNematicsScales,Yeomans2025-fe} and for material applications \cite{Harirchi2024-qa, sokolov_2025_SyntheticActiveLiquid,Wang2025-ww,Luo2026-oj,Lei2026-hs}.
In active nematics, constituent units determine their orientation at each point and generate singular points called topological defects where the orientation cannot be defined.
Specifically, topological defects are mathematically characterized by their winding numbers, known as topological charges.
When the cell alignment rotates by $2 \pi q$ counterclockwise around a defect, the defect has a topological charge of $q$, where $q$ is a nonzero half- or full-integer.
Because the symmetry of the alignment around a defect varies according to its charge, the spatial distribution of active stress becomes nonuniform in its vicinity, driving various biomechanical phenomena such as apoptosis \cite{saw_2017_TopologicalDefectsEpithelia}, rupture \cite{sonam_2023_MechanicalStressDrivenb}, aggregation \cite{kawaguchi_2017_TopologicalDefectsControl, kaiyrbekov_2023_MigrationDivisionCell}, calcium wave propagation \cite{Winterstrain2026-xq}, wound closure \cite{Andralojc2026-lh}, and layer formation \cite{copenhagen_2021_TopologicalDefectsPromote}.
In addition, other authors have recently reported that the geometries of cellular tissues influence the spatial positions of defects, around which morphological changes are triggered by nonuniform stress distributions; these changes include cell protrusion \cite{guillamat_2022_IntegerTopologicalDefects}, gastrulation \cite{Li2025-su}, lumen nucleation \cite{Guruciaga2026-hn}, and organogenesis in \textit{Hydra} \cite{maroudas-sacks_2021_TopologicalDefectsNematic, vafa_2022_ActiveNematicDefects, ravichandran_2025_TopologyChangesHydra}.
These biological phenomena occur on two-dimensional surfaces of tissues and are thus modeled as the dynamics of topological defects in two-dimensional geometries.
Therefore, investigating how the geometry of the domain determines the configurations of topological defects in two-dimensional geometries is important for understanding the biomechanical functions of cellular tissues.

Among the geometrical effects of cell culture domains, topological features such as the Euler characteristic play an important role in determining the topological charge of each defect.
When the cell alignment satisfies boundary anchoring conditions, where cells align parallel or perpendicular to all boundaries, experimental studies have revealed that the total topological charge generated within the domain is strictly equal to its Euler characteristic $\chi$ \cite{duclos_2017_TopologicalDefectsConfined, miyazako_2022_ExplicitCalculationMethod, ienaga_2023_GeometricConfinementGuides, guillamat_2022_IntegerTopologicalDefects, maroudas-sacks_2021_TopologicalDefectsNematic}.
For example, $\chi$ is equal to 1 for a two-dimensional simply connected domain, which implies that topological defects with positive charges must be generated.
Indeed, Duclos {\it et al.} \cite{duclos_2017_TopologicalDefectsConfined} first showed that two defects with charges of $+1/2$ ($+1/2$ defects) were stably formed when cells were cultured within small disks.
Other research groups have also confirmed this constraint on the total topological charge for other simply connected domains ($\chi = +1$) \cite{guillamat_2022_IntegerTopologicalDefects, ienaga_2023_GeometricConfinementGuides, miyazako_2024_PredictiveModelSpatialb} and closed surfaces ($\chi = +2$) \cite{maroudas-sacks_2021_TopologicalDefectsNematic,eckert_2025_CapturingNematicOrdera}.
Researchers have also theoretically predicted stable configurations of topological defects by minimizing the elastic energy originating from cell alignment, known as the Frank elastic energy \cite{duclos_2017_TopologicalDefectsConfined, miyazako_2022_ExplicitCalculationMethod}.
For an annular domain ($\chi = 0$), Miyazako {\it et al.} \cite{miyazako_2024_DefectPairsNematic} theoretically showed that the inner radius determines whether a pair of $+1/2$ and $-1/2$ defects forms or cells align circumferentially without defects.
Experimentally, several mammalian cells in annular domains exhibit circumferential alignment with chirality \cite{wan_2011_MicropatternedMammalianCells}, whereas pairs of $+1/2$ and $-1/2$ defects form near boundaries in a confined {\it fd}-virus \cite{garlea_2016_FiniteParticleSize}.

As discussed above, previous studies have mainly focused on domains with non-negative Euler characteristics and have successfully identified stable defect configurations through both experiments and theory.
By contrast, domains with negative Euler characteristics have not been fully investigated even though they have distinct properties regarding domain connectedness (i.e., the number of holes).
For a two-dimensional closed domain, the Euler characteristic $\chi$ is given by $1-h$, where $h$ is the number of holes in the domain.
When $h \geq 2$, the Euler characteristic becomes negative, which means that topological defects with negative charges must be formed in a closed domain with two or more holes.
Under these conditions, the interplay between defects and holes must be considered because it models the biomechanics of tissues containing internal obstacles, such as tumor growth \cite{argento_2025_ThreedimensionalTopologicalDefects} or the rupture of epithelial tissue \cite{sonam_2023_MechanicalStressDrivenb}.
Specifically, when the myocardial fiber orientation in short-axis slices of the human heart is observed by polarized-light microscopy \cite{jouk_2021_MyosinMyocardialMesh, auriau_2022_NematicChiralLiquid}, two isolated $-1/2$ defects are identified at the anterior and posterior junctions between the two ventricles \cite{auriau_2023_SoftMatterPhysicsProvides}, which can be viewed as nematic order in a triply connected domain. 
The two identified $-1/2$ defects appear in sequential slices and can thus be modeled as disclination lines running along the apico-basal axis in three-dimensional space \cite{auriau_2022_NematicChiralLiquid}.
Because the dynamics of disclination lines can be approximated as those of point defects in two-dimensional nematic order, understanding the behavior of defects in two-dimensional triply connected domains can provide valuable insights into the relationship between the disclination lines and three-dimensional heart mechanics \cite{kawahira_2026_TopologicalDefectsCoherentb}.

From a theoretical viewpoint, Miyoshi {\it et al.} \cite{miyoshi_2025_AnalyticalFormulasAlignment} used an explicit formula for nematic alignment to reveal multiple stable positions for two $-1/2$ defects in a specific triply connected domain.
However, how actual cell populations achieve stable defect configurations in the presence of internal obstacles remains unclear because the assumptions regarding the total charge and the number of defects do not necessarily hold in cell culture experiments.
To precisely model the mechanical interaction between obstacles and cellular tissues, it is necessary to elucidate the geometrical effects of obstacles on cell alignment and defects in two-dimensional systems through both experiments and theory.

The aim of the present study is to elucidate stable configurations of topological defects in triply connected domains, which represent the simplest geometry possessing a negative Euler characteristic.
To this end, we first present experimental results obtained by culturing C2C12 myoblast monolayers on microwells containing two circular obstacles, demonstrating that two $-1/2$ defects are predominantly observed when the internal obstacles are sufficiently large compared with the cell size.
To account for these experimental observations, we then theoretically analyze the configuration of the two $-1/2$ defects using the numerical framework proposed in our previous study \cite{miyoshi_2025_AnalyticalFormulasAlignment}.
Finally, we discuss the validity of the theoretical framework by comparing numerical results with the results of cell culture experiments.

\section{Cell culture experiments}
\label{sec:experiments}

\subsection{Defect pair annihilation in confined mouse myoblasts in triply-connected domains}

\begin{figure}[!h]
    \centering
    \includegraphics[
        width=\textwidth,
        trim=0 6.9cm 0 0,
        clip
    ]{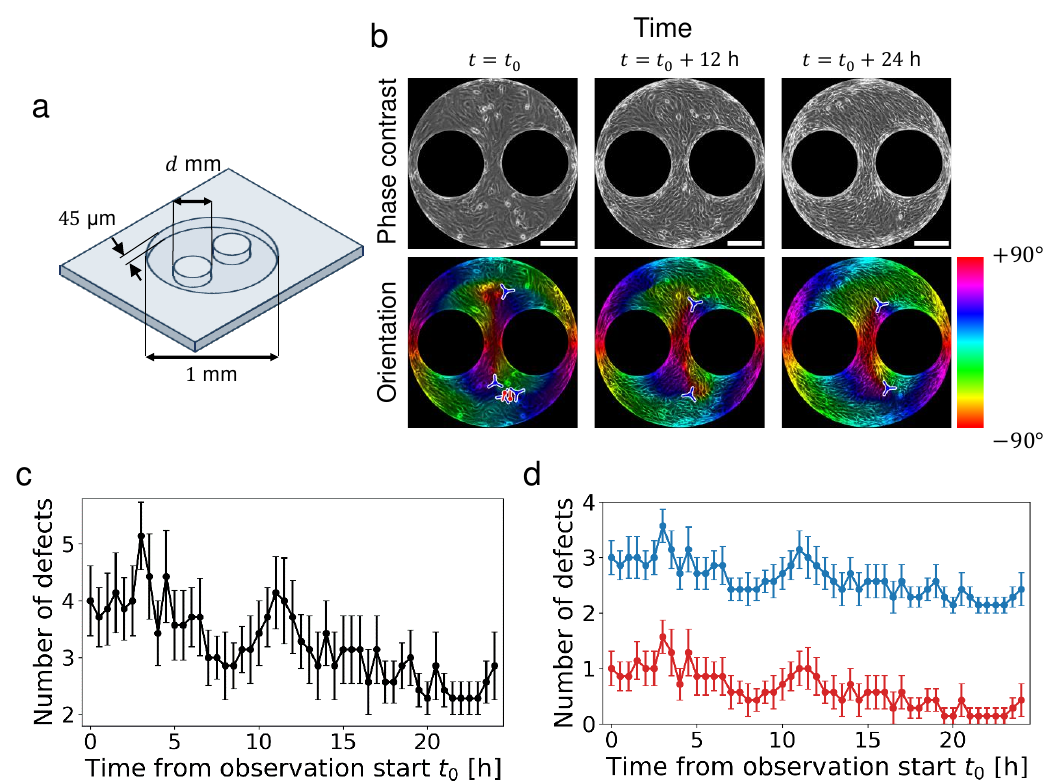}
    \caption{
        Fabricated PDMS geometries and time-lapse observation of cell cultures.
        (a) Schematic of the substrate geometry. The diameters of the outer and inner circles were $1.0$~mm and $d=0.1, 0.2, 0.3, 0.4$~mm, respectively.  The well depth was set to 45~\textmu m.
        (b) Time-lapse images of cell populations on Pattern 4 ($d=0.4$~mm). The time $t_0$ represents the start point of the observation (approximately 1 day after seeding). (Top) Phase contrast images. (Bottom) Color maps of the orientation field. Red and blue marks represent positions of $+1/2$ and $-1/2$ defects, respectively. Scale bars: 200~\textmu m.
        (c) Temporal dynamics of the mean number of defects in Pattern 4 ($n = 7$ independent substrates).
        Error bars represent the standard error of the mean (SEM).
        (d) Temporal dynamics of the mean number of $+1/2$ (red line) and $-1/2$ (blue line) defects in Pattern 4. 
    }
    \label{fig:fig1}
\end{figure}

We cultured C2C12 mouse myoblasts on polydimethylsiloxane (PDMS) substrates to examine how stable configurations of $-1/2$ defects are determined in triply connected domains.
To simplify the effects of geometry, we fabricated a 1\,mm-diameter circular hole containing two circular pillars of diameter $d$ symmetrically placed about the origin (Fig.~\ref{fig:fig1}a).
The diameter $d$ was set to 0.1, 0.2, 0.3, and 0.4\,mm to assess the dependence of stable defect configurations on the geometry. These four geometries are hereafter referred to as Patterns 1, 2, 3, and 4, respectively.

We first observed the dynamics of cell alignment and defect movement after the cells reached a confluent state, where they began to exhibit long-range nematic order.
To enhance the nematic order through boundary anchoring, we focused on Pattern 4 because it provided the smallest domain area among the four patterns.
We then started time-lapse imaging at 30-min intervals, beginning one day after cell seeding ($t = t_0$; Movie~S1, Fig.~\ref{fig:fig1}b).
At the initial state ($t = t_0$), the cells were aligned along all of the boundaries.
Concurrently, both $+1/2$ and $-1/2$ defects were observed within the domain.
As the observation progressed, a pair of $+1/2$ and $-1/2$ defects approached and annihilated each other. 
As a result, only the two $-1/2$ defects remained in the domain at $t = t_0 + 12\,\mathrm{h}$.
This defect configuration remained stable until $t = t_0 + 24\,\mathrm{h}$, approximately 2 days after seeding.

For statistical analysis of the defect dynamics, the number of defects at each time point was quantified and averaged over seven independent experiments (Fig.~\ref{fig:fig1}c,d).
At the initial state ($t = t_0$), the mean total number of defects was four: one  $+1/2$ defect and three $-1/2$ defects.
During the 24\,h period, the mean total number of defects decreased from four to two; specifically, the $+1/2$ defect count decreased to zero, whereas the $-1/2$ defect count decreased to two.
These results indicate that a pair of $+1/2$ and $-1/2$ defects annihilated during this 24-h period.
Throughout this defect annihilation process, the mean total topological charge in the domain remained $-1$, whereas its standard deviation decreased over time.
Consequently, the topological constraint imposed by the surface anchoring condition tended to be more strictly satisfied over the course of the observation.

\subsection{Geometrical dependence on defect configuration statistics}

\begin{figure}[t]
    \centering
    \includegraphics[
        width=\textwidth,
        trim=0 3.25cm 0 0,
        clip
    ]{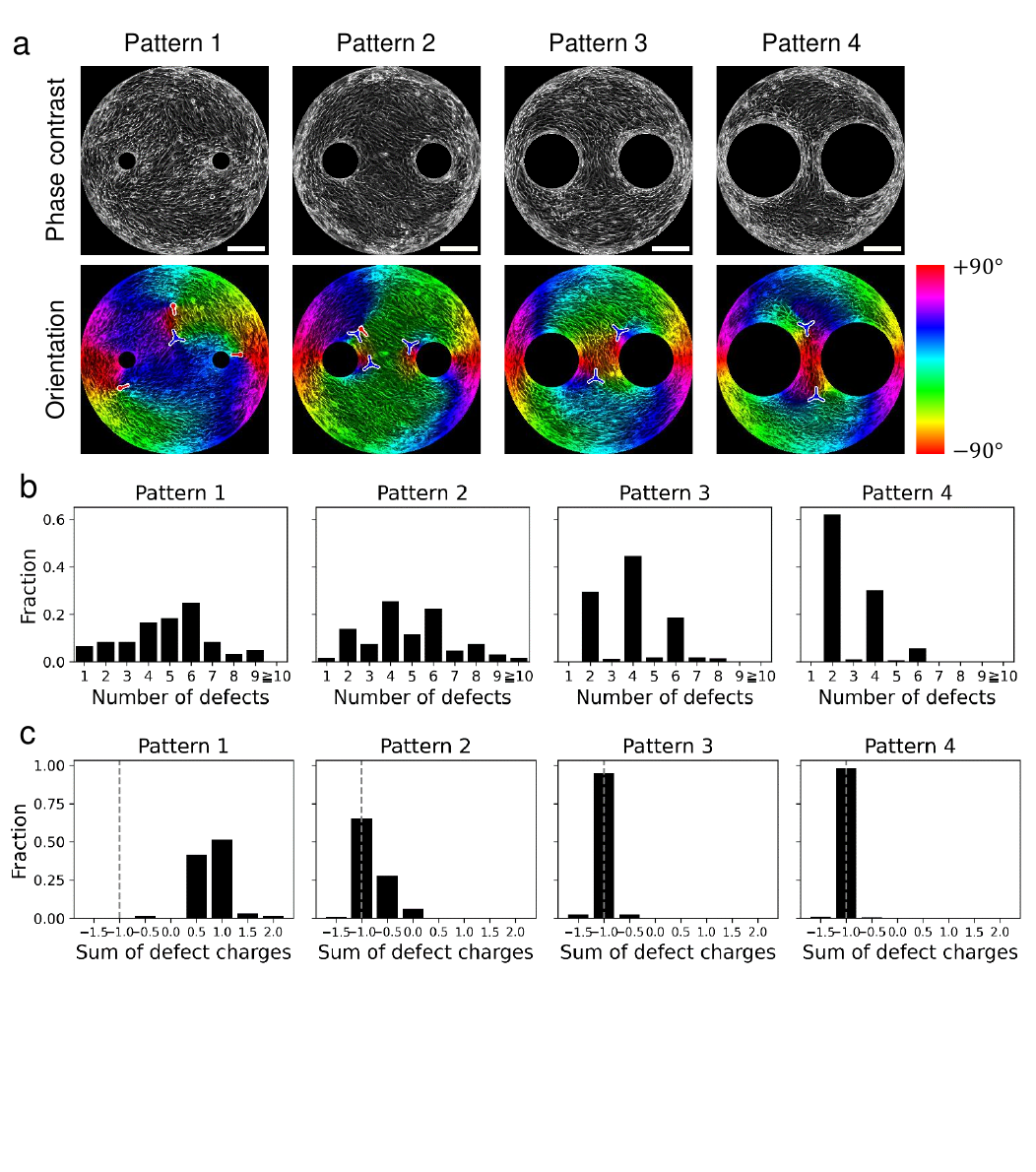}
    \caption{
        Statistics for the defect number and total topological charge for each pattern.
        (a) Cell alignment 2--3 days after cell seeding . (Top) Phase contrast images. (Bottom) Orientation field. Red and blue marks represent $+1/2$ and $-1/2$ defects, respectively. Scale bars: 200~\textmu m.
        (b) Distribution of the total number of defects per substrate for Patterns 1--4. Note that samples whose defect count was ten or greater were merged into one bin ($\geq 10$). 
        (c) Distribution of the total topological charge per substrate for Patterns 1--4. Dashed vertical lines at $-1.0$ represent the theoretical anchoring constraint.
        Number of images: 60 (Pattern 1), 129 (Pattern 2), 327 (Pattern 3), 314 (Pattern 4).
    }
    \label{fig:fig2}
\end{figure}

We next investigated the dependence of defect configurations on geometry by culturing cells on Patterns 1--4 and counting the defects 2--3 days after cell seeding (Fig.~\ref{fig:fig2}a).
As the pillar diameter $d$ decreased, the total number of defects increased, accompanied by the generation of additional $+1/2$ defects.
Notably, for Pattern 1, the cells failed to align along the inner pillars in several cases, indicating that the surface anchoring condition was not satisfied.

Statistical analysis of the defect counts for Pattern 4 revealed that configurations with two defects were the most frequently observed, followed by those with four and six defects (Fig.~\ref{fig:fig2}b).
For Pattern 3, configurations with four defects were the most predominant, followed by those with two and six defects.
Although the defect number distributions differed between Patterns 3 and 4, the proportion of configurations with a total charge of $-1$ exceeded $95\%$ in both cases (Fig.~\ref{fig:fig2}c).
Therefore, although the topological charge constraint was satisfied for both patterns, defect annihilation tended to remain incomplete in Pattern 3.

The defect counts were consistently even for Patterns 3 and 4, whereas configurations with an odd number of defects were also frequently observed for Pattern 2, although configurations with four defects remained the most predominant. Correspondingly, configurations with total topological charges of $-0.5$ and $0$ were observed (Fig.~\ref{fig:fig2}c). Furthermore, no clear peak was observed in the defect number distribution for Pattern 1, whereas the total topological charge distribution exhibited peaks at $+0.5$ and $+1.0$. This deviation from the total topological charge constraint implies that the surface anchoring condition failed to hold for Patterns 1 and 2.

\clearpage

\subsection{Spatial statistical analysis of defect distributions}
\label{subsec:spatial analysis}

\begin{figure}[!h]
    \centering
    \includegraphics[
        width=\textwidth,
        trim=0 13.25cm 0 0,
        clip
    ]{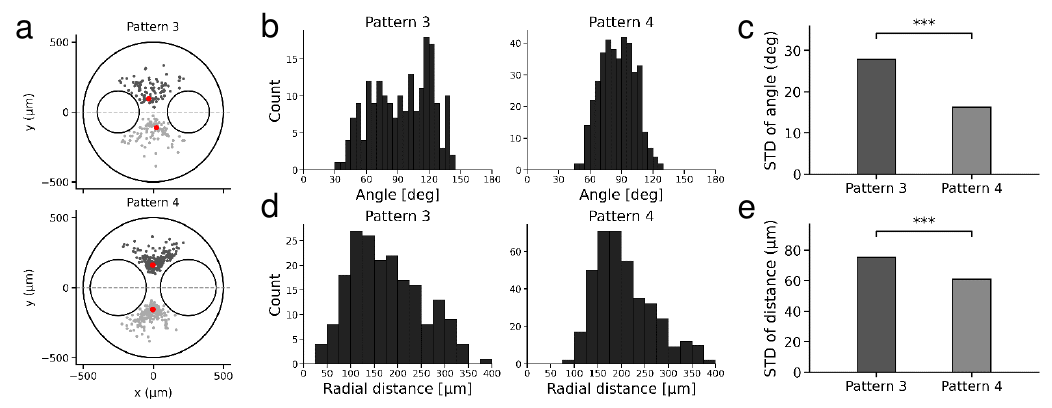}
    \caption{
        Spatial statistics for defects configurations.
        (a) Overlay of defect positions from images with two $-1/2$ defects. Number of images: 97 (Pattern 3), 195 (Pattern 4). Dark- and light-gray markers represent defects in the upper and lower half-planes, respectively. Red dots indicate the most likely defect positions in each half-plane, estimated as the peak of the defect distributions approximated by KDE.
        (b, d) Histograms of (b) the angle of defect positions from the positive horizontal axis and (d) the radial distance of defect positions from the origin. Note that lower half-plane defects are reflected with respect to the origin so that the angles are defined in $[0^\circ, 180^\circ)$.
        (c, e) Standard deviations (STDs) of the distributions in (b) and (d), respectively; *** indicates that the $p$-value is less than 0.001.
    }
    \label{fig:fig3}
\end{figure}

To further examine the geometrical effects on defect positions, we focused on the spatial distributions of defect configurations containing two $-1/2$ defects in Patterns 3 and 4, where the topological charge constraint was satisfied (Fig.~\ref{fig:fig3}a).
In all cases, one $-1/2$ defect was located on the top side, whereas the other was positioned on the bottom side.
The spatial defect distributions were approximated by a two-dimensional Gaussian kernel density estimate (KDE; see Appendix~\ref{app:image}), which suggested that the most probable position of each cluster was near the central vertical axis (red points in Fig.~\ref{fig:fig3}a).
On the basis of these spatial distributions, we constructed histograms of the defect angles and their radial distances from the origin (Fig.~\ref{fig:fig3}b, d).
In Pattern 4, both the angular distribution and radial distributions were unimodal.
By contrast, in Pattern 3, no peak was observed around $90^{\circ}$, whereas distinct peaks were observed at $60^{\circ}$--$75^{\circ}$ and $105^{\circ}$--$125^{\circ}$.
For the radial distributions, the peak in Pattern 3 was shifted by approximately 50~\textmu m toward the origin compared with that in Pattern 4.
These results demonstrate that the defect distributions shifted away from the central vertical axis and approached the origin in Pattern 3.
In addition, to compare the variability of the defect positions between the two patterns, we applied Levene's test to the angle and radial distance distributions. The results showed that the standard deviations for Pattern 3 were significantly greater than those for Pattern 4 for both the angle and the radial distance (Fig.~\ref{fig:fig3}c, e).
These results indicate that the defect positions were more tightly localized in Pattern 4 than in Pattern 3.

\begin{figure}[!h]
    \centering
    \includegraphics[
        width=\textwidth,
        trim=0 3.5cm 0 0,
        clip
    ]{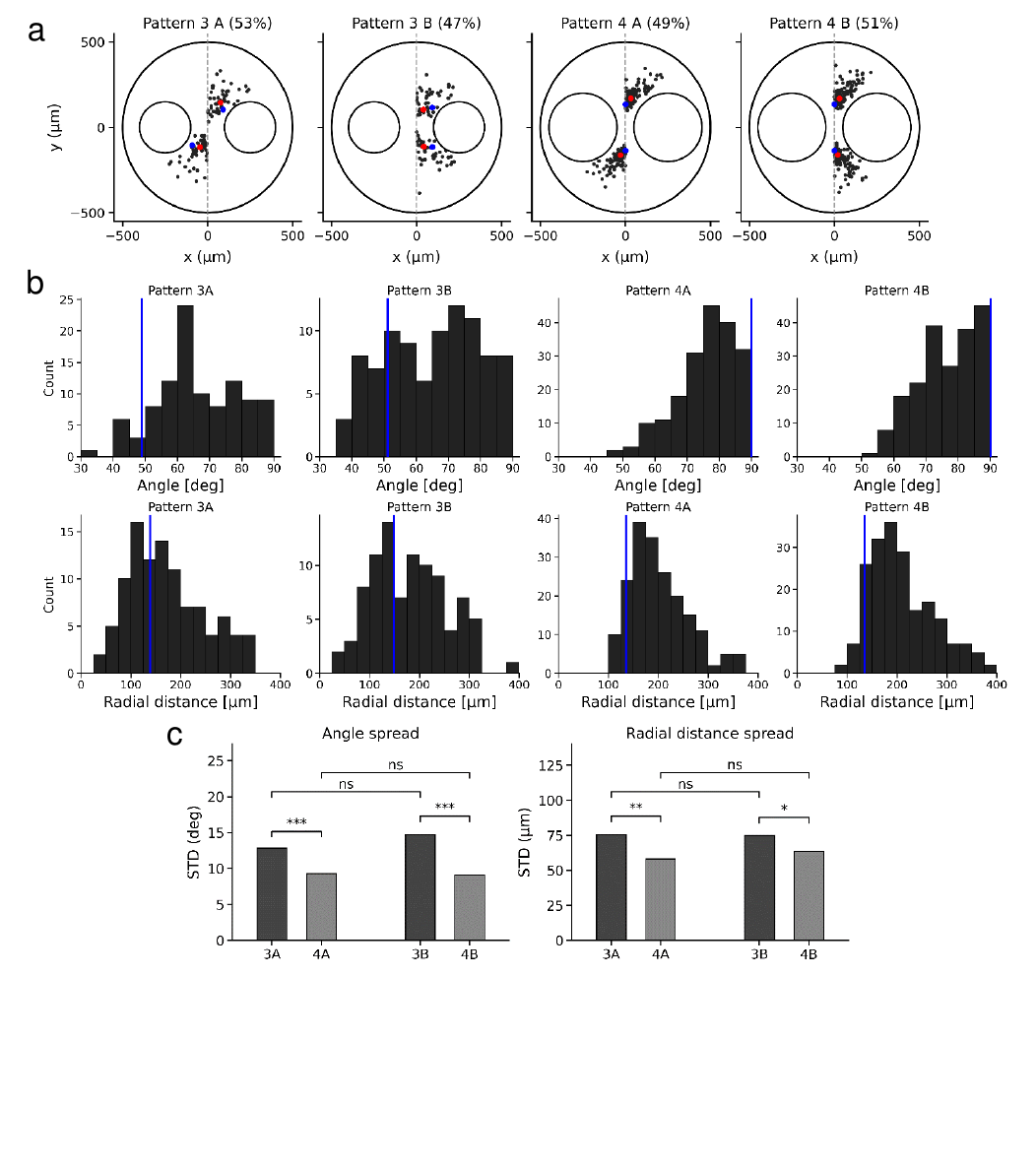}
    \caption{
        Statistical analysis of defect configurations based on their symmetry.
        (a) Overlay of defect positions from images with two $-1/2$ defects for Groups A and B. Each pair was reflected; thus, the upper defect lies in the first quadrant. Red dots indicate the most probable defect positions estimated by KDE, as in Fig.~\ref{fig:fig3}a. Blue dots indicate theoretically predicted stable defect configuration described in Section \ref{sec:numerical}. Percentages: fraction of two-defect images. Dashed lines represent the horizontal axes. 
        (b) Histograms of the polar angle (top) and radial distance from origin (bottom) of defects. Blue lines: theoretically predicted stable defect configuration described in Section \ref{sec:numerical}.
        (c) STDs of the angle (left) and radial distance (right) distributions in (b) for each symmetry class. *, **, and *** indicate that the $p$-value is less than 0.05, 0.01, and 0.001, respectively; ns indicates no significant difference.
    }
    \label{fig:fig3b}
\end{figure}

We also evaluated the symmetry of the defect configurations.
To this end, the geometries were divided into four quadrants by the central horizontal and vertical axes.
The configurations were then classified into two groups: Group A, where the two $-1/2$ defects were located either in the first and third quadrants or in the second and fourth quadrants, and Group B, where the defects were located either in the first and fourth quadrants or in the second and third quadrants (Fig.~\ref{fig:fig3b}a).
In addition, for the statistical analysis, the configurations with one $-1/2$ defect in the second quadrant were flipped horizontally with respect to the central vertical axis.
The results showed that configurations of Groups A and B were observed at comparable frequencies in both Patterns 3 and 4.
Histograms of the defect positions were then constructed for each symmetry group (Fig.~\ref{fig:fig3b}b).
For both Groups A and B, the peaks shifted toward smaller angles with decreasing inner pillar diameter $d$, consistent with the results shown in Fig.~\ref{fig:fig3}b.
Similarly, Patterns 3A and 3B exhibited larger standard deviations in the defect angle than Patterns 4A and 4B (Fig.~\ref{fig:fig3b}c).

In this section, we demonstrated that the stable defect configurations with two $-1/2$ defects were reproducibly observed for Patterns 3 and 4. In these patterns, the topological constraints for boundary anchoring and topological charges were satisfied because of the highly organized nematic order of the myoblasts. In particular, two distinct defect configurations were observed in Pattern 3, whereas the defects tended to be located around the vertical axis in Pattern 4. These results suggest that the diameter of the two inner circles strongly affects both the stable defect configurations and the symmetry of cell alignment. In the following sections, we present numerical studies to explain the geometrical effects of the inner obstacles on the stable defect configurations.

\clearpage %

\section{Theoretical framework of stable defect configurations}
\label{sec:theory}

To theoretically explain the stable configurations observed in the cell culture experiments, we introduce a theoretical framework for cell alignment based on the physics of nematic liquid crystals. In this section, we summarize the mathematical methods for predicting the nematic cell alignment in triply connected domains and explain the numerical procedure for detecting stable defect configurations, which is based on the formulas described in our previous study \cite{miyoshi_2025_AnalyticalFormulasAlignment}.

\begin{figure}[h]
\centering
  \includegraphics[
        width=\textwidth,
        trim=0 11.8cm 0 0,
        clip
    ]{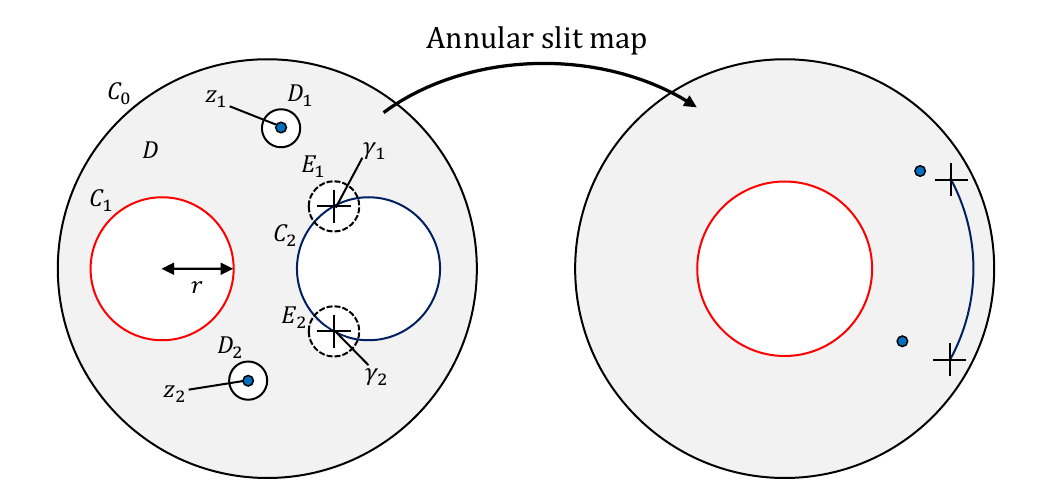}
  \caption{Geometry of a triply connected domain $D$. The domain contains two internal holes of equal radius $r$ and hosts two $-1/2$ topological defects (blue dots) located at $z_1$ and $z_2$. $D_k$ and $E_k$ ($k=1,2$) are excluded for the calculation of the Frank elastic energy.}
  \label{fig:geometry}
\end{figure}

We consider a triply connected domain $D$ identical to that used in the experiments (Fig.~\ref{fig:geometry}); the domain contains an outer circular boundary $C_0$ and two identical internal circles $C_1$ and $C_2$ arranged symmetrically.
Throughout the theoretical analysis, all lengths are normalized by the outer-circle radius; thus, the outer boundary has a unit radius, and each inner circle has a radius $r$ centered at a distance $\delta$ from the origin on the real axis. The outer and inner circles can be defined as follows:
\begin{eqnarray*}
    C_0 &=& \{(\cos\theta,\,\sin\theta)\,|\,0\leq\theta<2\pi\}, \\
    C_1 &=& \{(r\cos\theta+\delta,\,r\sin\theta)\,|\,0\leq\theta<2\pi\}, \\
    C_2 &=& \{(r\cos\theta-\delta,\,r\sin\theta)\,|\,0\leq\theta<2\pi\}.
\end{eqnarray*}
Under these assumptions and definitions, the four experimental conditions described in Section~\ref{sec:experiments} correspond to $r = 0.1, 0.2, 0.3$, and $0.4$ with $\delta = 0.5$. 
In addition, tangential anchoring on all boundaries is assumed to reflect the experimental results for Patterns 3 and 4. This boundary condition fixes the total topological charge in this domain to $-1$, which necessitates the presence of two $-1/2$ defects. Consequently, we fix the number of defects to two and focus on their spatial arrangement within the domain.

\begin{figure*}
  \centering
  \includegraphics[
        width=\textwidth,
        trim=0 13.5cm 0 0,
        clip
    ]{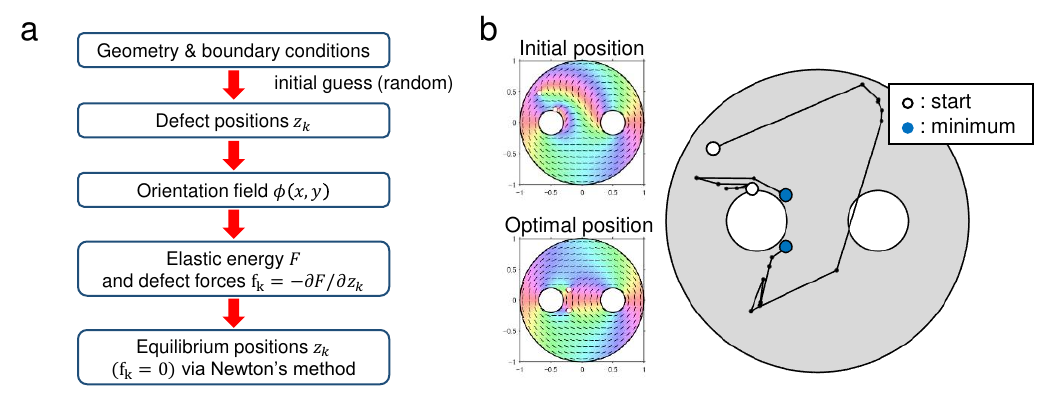}
  \caption{
    (a) Overview of the theoretical framework.
    For a given geometry and defect positions $z_1$ and $z_2$, the orientation field $\phi(x,y)$ is constructed analytically on the basis of Equation ~\eqref{eq:fDzeta} .
    The Frank elastic free energy and the elastic forces acting on defects are then evaluated, and defect positions are iteratively updated until the Frank elastic energy is minimized.
    (b) Representative example of the minimization of the Frank elastic energy via Newton's method.
    Starting from a random initial configuration (white dots), the defect positions converge to a critical point where $\mathrm{f}_k = 0$ (blue dots).
  }
  \label{fig:overview}
\end{figure*}

On the basis of the above assumptions, the theoretical calculations proceed in three steps to predict stable defect configurations, as outlined in Fig.~\ref{fig:overview}a.
First, we place two $-1/2$ defects within domain $D$ and analytically construct a holomorphic function that describes the corresponding orientation field (Section~\ref{subsec:orientation_field}). We then evaluate the Frank elastic free energy and the elastic forces acting on the defects using the constructed function. Finally, we determine the stable defect configurations by iteratively updating defect positions until the Frank elastic energy is minimized. (Section~\ref{subsec:energy_optimization}).

\clearpage

\subsection{Holomorphic expression of nematic orientation field}
\label{subsec:orientation_field}
Let us consider the alignment of cells at position $(x,y)$ within the domain $D$, which can be physically modeled as a two-dimensional nematic director field
$\mathbf{n}(x,y)=(\cos\phi(x,y),\sin\phi(x,y))^\mathrm{T}$ by neglecting the finite size of cells.
Because the nematic liquid crystals are apolar, the alignment angle $\phi(x,y)$ is defined on $[-\pi/2,\pi/2)$, where $\phi + m\pi$ is identified with $\phi$ for any integer $m$.

Miyoshi {\it et al.}\cite{miyoshi_2025_AnalyticalFormulasAlignment} derived an explicit formula for the orientation field in a multiply connected domain with an arbitrary number of holes and possessing defects of arbitrary charges in arbitrary positions.
Here, we apply their general formula to the specific triply connected domain $D$ with two $-1/2$ defects (Fig.~\ref{fig:geometry}).
The orientation field $\phi(x,y)$ in $D$ is characterized by the following conditions:

\begin{enumerate}
\item[(i)] The alignment angle $\phi(x,y)$ satisfies $\nabla^2 \phi(x,y)=0$ except at the locations of the defects.

\item[(ii)] Each defect carries a topological charge of $-1/2$, such that the winding number around
each defect satisfies
\begin{equation*}
\oint_{\Gamma_k} \frac{\partial \phi}{\partial s}\,ds = -\pi,
\end{equation*}
where $\Gamma_k$ is a Jordan curve that contains only the $k$-th defect ($k=1,2$). 

\item[(iii)] The alignment angle satisfies the following boundary conditions for tangential anchoring:
\begin{eqnarray*}
    \phi(\cos\theta,\sin\theta)&=& \theta-\frac{\pi}{2},\\
    \phi(r\cos\theta+\delta,r\sin\theta)&=& \theta-\frac{\pi}{2}+N_1\pi, \\
    \phi(r\cos\theta-\delta,r\sin\theta)&=& \theta-\frac{\pi}{2}+N_2\pi,
\end{eqnarray*}
\end{enumerate}
where $N_m$ ($m=1,2$) is an integer specifying additional half-rotations imposed around $C_m$.
Considering the experimental results presented in Section~\ref{sec:experiments}, we set $N_1 = N_2 = 0$ to exclude additional rotations in the following analysis.

Considering Condition (i), holomorphic functions can be applied to express the orientation field $\phi(x,y)$. To this end, we identify the two-dimensional coordinates $(x,y)$ with a complex number $z=x+iy$ and define the position of the $k$-th defect ($k=1,2$) as $z_k=x_k+iy_k$. The alignment angle $\phi(x,y)$ satisfying Conditions (i) and (ii) can then be expressed as the real part of the holomorphic function $f(z;z_1,z_2):=\frac{i}{2}(\log(z-z_1)+\log(z-z_2))+h(z)$, where $h(z)$ is a holomorphic function on $D$ determined to satisfy Condition (iii).
For the domain $D$, the holomorphic function $f(z;z_1,z_2)$ whose real part satisfies Conditions (i)--(iii) is explicitly obtained via a conformal mapping from $D$ to an annular-slit domain, which maps the boundaries $C_0$, $C_1$, and $C_2$ onto the outer circle, the inner circle, and a circular slit, respectively (Fig.~\ref{fig:geometry}). We denote the preimages of the two end points of the slit as $\gamma_1$ and $\gamma_2$. Using the Schottky--Klein prime function~\cite{crowdy_2020_SolvingProblemsMultiply}, we obtain the holomorphic function $f(z;z_1,z_2)$ analytically by
\begin{equation}
\begin{aligned}
f(z;z_1,z_2)=-i\pi
\sum_{j=1}^{2}
\mathcal{W}\!\left(
  z;\gamma_{j},z_j
\right)
+
\sum_{m=1}^{2} i
\left(
  c_m - \pi N_m
\right)
\hat{v}_m(z)
+
i \log \!\left( -v_1'(z) \right),
\end{aligned}
\label{eq:fDzeta}
\end{equation}
where $' = \frac{\partial}{\partial z}$, $\mathcal W(z;\alpha,\beta)$ denotes the logarithm of a ratio of Schottky--Klein
prime functions and $v_m(z)$ and $\hat v_m(z)$ are analytic continuations of the harmonic measures. The constants $c_m$ ($m=1,2$) are defined as
\begin{eqnarray*}
    c_m = -\pi\sum_{k=1}^2 \mathrm{Re}[v_m(z_k)]+\sum_{j=1}^2 \mathrm{Re}[v_m(\gamma_j)].
\end{eqnarray*}
In Equation \eqref{eq:fDzeta}, the first term represents the alignment change around each defect, whereas the second and third terms serve as complementary terms to satisfy tangential anchoring condition (iii).
We refer the reader to Ref.~\cite{miyoshi_2025_AnalyticalFormulasAlignment} for detailed definitions of the functions $\mathcal W(z;\alpha,\beta)$, $v_m(z)$, and $\hat{v}_m(z)$ and the derivation of Equation \eqref{eq:fDzeta}.

\subsection{Minimization of the Frank elastic energy}
\label{subsec:energy_optimization}
On the basis of the holomorphic function $f(z;z_1,z_2)$, the Frank elastic energy associated with the orientation field $\phi(x,y)$ can be calculated to evaluate the stability of each defect configuration. Notably, the holomorphic approach enables us to represent the Frank elastic energy as the following line integral form:
\begin{eqnarray}
F(z_1,z_2)
&=&
\frac{K}{2}
\iint_{\tilde{D}}\left|\frac{\partial f(z;z_1,z_2)}{\partial z}\right|^2 dxdy \nonumber \\
&=&
\frac{K}{2}\cdot \frac{1}{2i}\oint_{\partial\tilde{D}}\overline{ f(z;z_1,z_2)}\frac{\partial f(z;z_1,z_2)}{\partial z}dz,
\label{eq:frank_energy}
\end{eqnarray}
where $K$ is the Frank elastic constant under one constant approximation and $\tilde{D}=D\setminus(D_1(\epsilon_1)\cup D_2(\epsilon_1) \cup (E_1(\epsilon_2) \cap D) \cup (E_2(\epsilon_2) \cap D))$. 
$D_k(\epsilon_1)$ and $E_k(\epsilon_2)$ ($k=1,2$) are small circles centered at $z_k$ and $\gamma_k$ with radii of $\epsilon_1$ and $\epsilon_2$, respectively. When performing the numerical calculation of the line integral in Equation \eqref{eq:frank_energy}, we set $\epsilon_1$ and $\epsilon_2$ to $10^{-3}$ and $10^{-2}$, respectively.
To avoid discontinuities arising from logarithmic branch cuts in the holomorphic function $f(z;z_1,z_2)$, we chose an integration contour that does not intersect these cuts.

According to the cell culture experiments in existing studies \cite{duclos_2017_TopologicalDefectsConfined, miyazako_2024_PredictiveModelSpatialb}, the cell alignment tends to rearrange itself to minimize the elastic energy. Because the elastic energy \eqref{eq:frank_energy} is a function of $z_1$ and $z_2$, the Frank elastic energy acts as an effective potential for the defect positions, which enables us to consider the force acting on each defect during the minimization process. By minimizing the elastic energy with respect to the defect positions, steady defect configurations can be achieved, where all of the forces acting on defects vanish. Let $\mathrm{f}_k=\mathrm{f}_{kx}-i\mathrm{f}_{ky}$ be the force acting on the $k$-th defect whose $x$ and $y$ components correspond to $\mathrm{f}_{kx}$ and $\mathrm{f}_{ky}$, respectively. According to Miyoshi {\it et al.} \cite{miyoshi_2024_FreeEnergyFormulaea}, the force acting on the $k$-th defect is analytically derived as
\begin{equation}
\mathrm{f}_k
=
- \frac{\partial F(z_1,z_2)}{\partial z_k}=-\frac{i\pi K}{2}
\left.
\frac{\partial \hat f_k(z)}{\partial z}
\right|_{z=z_k}
\label{eq:defect_force},
\end{equation}
where
\begin{equation}
\hat f_k(z)
=
f(z;z_1,z_2)
-
\frac{i}{2}\log(z-z_k).
\end{equation}
This explicit expression enables elastic forces to be directly evaluated without numerical differentiation of the energy \eqref{eq:frank_energy}.
To identify energetically stable defect configurations where $\mathrm{f}_k$ becomes zero for all $k$, we solve $\mathrm{f}_k = 0$ by Newton's method (Fig.~\ref{fig:overview}b). Details of the numerical procedure are given in Appendix~\ref{app:optimization}.

\section{Numerical Experiments}
\label{sec:numerical}

This section is dedicated to demonstrating how stable defect configurations containing two $-1/2$ defects are determined on the basis of the Frank elastic energy described in Section ~\ref{sec:theory}.
Let $z_1$ and $z_2$ be the positions of the two defects.
To simplify the calculations and discussion, we consider the symmetry of the defect configurations (Groups A and B defined in Section ~\ref{subsec:spatial analysis}) and assume two idealized defect configurations: (i) $z_2=-z_1$ and (ii) $z_2=\bar{z}_1$.
Under these configurations, the elastic energy can be expressed as a function of the single variable $z_1$, which enables a two-dimensional representation of the energy landscape and subsequent local stability analysis around critical points.

\subsection{Geometrical effects on energy landscape}
\label{subsec:identification}

\begin{figure}[!h]
    \centering
    \includegraphics[
        width=\textwidth,
        trim=0 11cm 0 0,
        clip
    ]{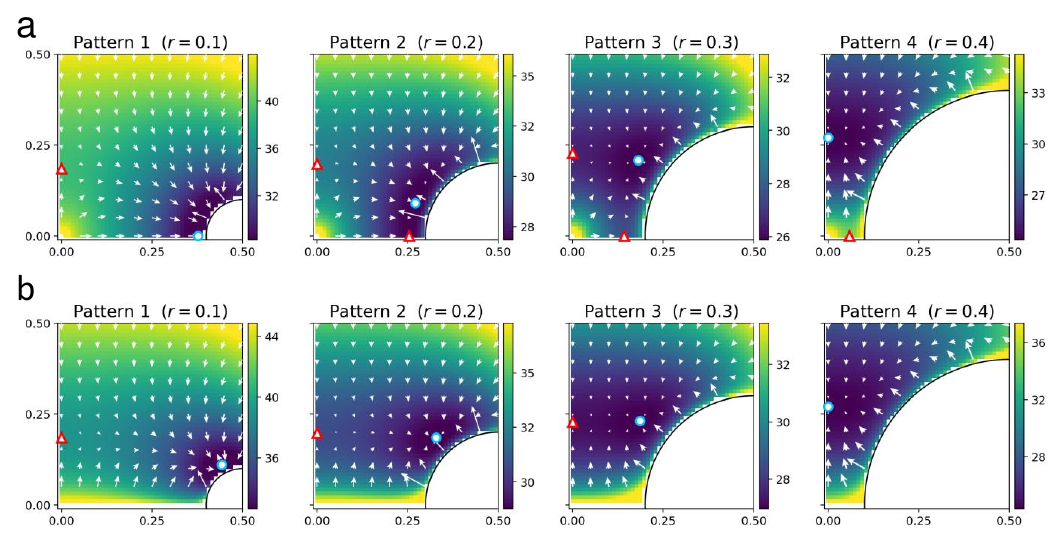}
    \caption{
        Frank elastic energy landscape and defect force field as a function of defect pair configuration $(z_1,z_2)$ for each pattern.
        (a) Configuration (i): $z_2=-z_1$.
        (b) Configuration (ii): $z_2=\bar{z}_1$.
        In each panel, the inner circle radius is $r$ and the position of the first defect $z_1$ is located in the first quadrant.
        The color map represents the Frank elastic energy $F$ of the system. White arrows indicate the force acting on the first defect $\mathrm{f}_1=-\partial F/\partial z_1$.
        Blue circles and red triangles denote energy minima and saddle points, respectively.
    }
    \label{fig:fig7}
\end{figure}

We first considered the four geometries corresponding to the cell culture experiments (Fig.~\ref{fig:fig1}) and constructed the energy landscape for each geometry under configuration (i) $z_2=-z_1$.
We then identified the positions of the energy critical points and examined their stability from the gradient around each critical point (Fig.~\ref{fig:fig7}a).
For Pattern 4, two critical points existed on the $x$- and $y$-axes, indicated by the red triangle and blue circle in Fig.~\ref{fig:fig7}a, respectively.
Local linear stability analysis revealed that the former was a saddle point, whereas the latter was a minimum.
By contrast, for Patterns 3 and 2, the critical point on the $x$-axis remained a saddle, whereas the critical point on the $y$-axis transformed to a saddle point.
Furthermore, a new energy minimum emerged off the $x$- and $y$-axes, which had no counterpart in Pattern 4.
For Pattern 1, the critical points on the $x$- and $y$-axes became a minimum and remained a saddle, respectively, whereas the off-axis minima found for Patterns 2 and 3 vanished.
These findings demonstrate that the domain geometry governs the local stability of the defect configurations.

The same analysis was performed under configuration (ii) $z_2=\bar{z}_1$ to identify the energy critical points and examine their stability (Fig.~\ref{fig:fig7}b).
In this case, a defect on the $x$-axis yields $z_2 = z_1$, which violates the premise that each defect carries a $-1/2$ charge; therefore, the critical points on the $x$-axis found under configuration (i) do not exist.
For the critical points on the $y$-axis, the results of the local stability analysis were consistent with those obtained under configuration (i), yielding a minimum for Pattern 4 and a saddle for Patterns 3, 2, and 1.
Similarly, off-axis minima were found for Patterns 3 and 2, as in the case of configuration (i); however, a minimum was also detected for Pattern 1, which had been absent under configuration (i).

The above analysis revealed that a unique energy minimum was found for each combination of the two assumptions and inner circle radius.
Specifically, the stable configurations of the defect pair at these energy minima can be classified into four distinct cases: (A) both defects lying on the $y$-axis symmetrically about the origin, (B) both on the $x$-axis in the same manner, (C) off-axis with point symmetry under configuration (i), and (D) off-axis with reflection symmetry about the $x$-axis under configuration (ii) (Fig.~\ref{fig:fig9}a).

\subsection{Bifurcation of energy minimum because of geometrical confinement}
\label{subsec:global_minimum}

We further evaluated which of the four configurations is the most energetically stable as a function of the inner circle radius $r$.
To this end, we calculated the Frank elastic energies for Cases A and B by varying $r$.
For Cases C and D, we numerically searched for the energy minima using Newton's method (see Appendix~\ref{app:optimization} for details).
As a result, stable energy minima were obtained for $0.18 \leq r \leq 0.38$ in Case C and for $r \leq 0.37$ in Case D.

\begin{figure}[!h]
    \centering
    \includegraphics[
        width=\textwidth,
        trim=0 5.2cm 0 0,
        clip
    ]{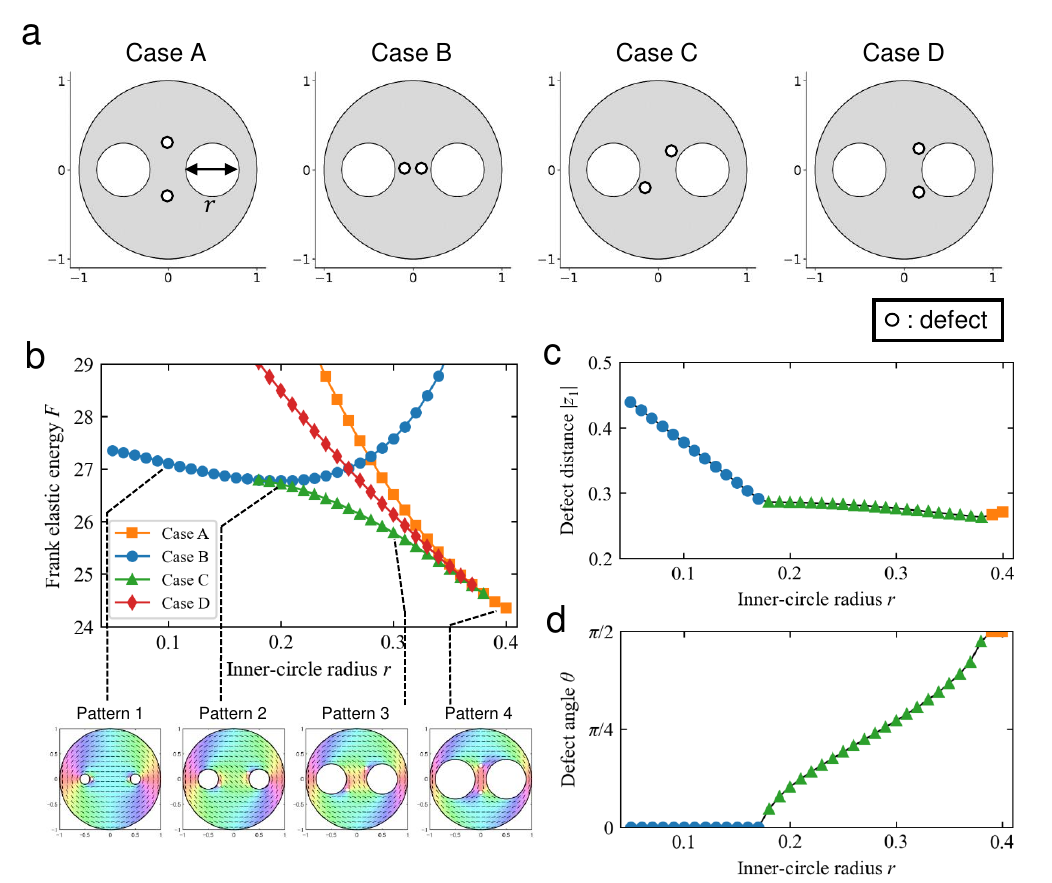}
    \caption{
        Calculation of the Frank elastic energy for searching for the most stable defect configurations.
        (a) Defect configurations for the energy analysis.
        (b) Calculations of the Frank elastic energy for various inner circle radius $r$. Note that comparisons are valid only among cases with the same $r$ because the domain area varies with $r$.
        (c, d) Dependence of the defect positions on the inner circle radius $r$ that achieve the minimized elastic energy. (c) Radial distance of the defects from the origin. (d) Angle of the defect positions from the positive horizontal axis.
        }
    \label{fig:fig9}
\end{figure}

According to the calculated energy (Fig.~\ref{fig:fig9}b), Case B was the most stable when $r < 0.18$ and the most stable configuration switched from Case B to Case C at $r=0.18$.
For $0.18 \leq r \leq 0.37$, the energy for Case D was always higher than that for Case C, meaning that Case D never achieved the global minimum for the elastic energy.
Finally, Case A became the most stable at $r=0.39$.
Notably, the minimum of the elastic energy was continuously changed at these bifurcation points ($r=0.18, 0.39$).

Fig.~\ref{fig:fig9}c and ~\ref{fig:fig9}d show the radial position and the angle from the horizontal axis of the minimizer $z_1$.
The radial position monotonically decreased for $r < 0.39$ and then increased after the second bifurcation point ($r=0.39$).
The defect angle monotonically increased from $0$ to $+\pi/2$ for $0.18 \leq r \leq 0.38$.
These results indicate that the most stable defect configurations and their elastic energy changed continuously as a function of the inner radius $r$.

Finally, we compared the above theoretical results with the experimentally observed defect positions reported in Section~\ref{sec:experiments}. In particular, we estimated the peak of the spatial defect distributions presented in (Fig.~\ref{fig:fig3b}a, red points) by the kernel density method (see Appendix~\ref{app:image} for details) and compared the estimated peak with the theoretical minima of the Frank elastic energy (Fig.~\ref{fig:fig3b}a, blue points). The peak of Pattern 3A and 3B agreed with the theoretical value for Case~C and ~D, respectively, and the peaks of Patterns 4A and 4B agreed with the theoretical value for Case~A. The gaps between the theoretical and experimental values were within 60~\textmu m in all cases, which is smaller than the typical long-axis length of a single C2C12 myoblast ($\approx$ 80~\textmu m) \cite{sheets_2013_ShapedependentCellMigration}, meaning that the proposed theoretical framework has sufficient predictive capability to explain the stable defect configurations observed in the cell culture experiments.

\clearpage

\section{Discussion}
\label{sec:discussion}

Our experimental and theoretical investigations revealed that nematic cell alignment in a triply connected domain stably generates two $-1/2$ defects when the cells are perfectly aligned along all boundaries (Patterns 3 and 4 in Section ~\ref{sec:experiments}), which corresponds to the main assumption for the theoretical analysis. This result demonstrates that the sign of the net topological charge is determined by the number of holes, which is consistently predicted by the Euler characteristic of the domain. In addition, the stable defect configurations observed in the experiments were accurately predicted by our proposed theoretical framework, which was built upon nematic liquid crystal theory and complex analysis. Interestingly, our theoretical investigations suggest that stable defect configurations can become off-axis symmetric (Cases C and D in Section ~\ref{sec:numerical}) when the radii of the two inner circles are intermediate (Fig.~\ref{fig:fig7}), which was actually observed in the experiments (Pattern 3 in Section ~\ref{sec:experiments}). Thus, the symmetry of the cell alignment is strongly affected by the size of inner obstacles in multiply connected domains, which was not considered for simply connected domains in previous studies \cite{duclos_2017_TopologicalDefectsConfined,saw_2017_TopologicalDefectsEpithelia,ienaga_2023_GeometricConfinementGuides,miyazako_2024_PredictiveModelSpatialb}. Given that the cell alignment and defect configurations dictate the direction of active stress that drives the development of organs and muscle contraction, changes in cell alignment symmetry could be critical for determining organ shape. Therefore, the proposed theoretical framework will be a promising foundation for further theoretical analyses to study relationships between cell alignment and tissue geometries.

By contrast, Patterns~1 and 2 in the experiments did not predominantly exhibit stable defect configurations with two $-1/2$ defects, which is attributable to several potential factors. First, the cells failed to align tangentially along the inner boundaries in Pattern 1 (Fig.~\ref{fig:fig1}), possibly because tangential anchoring could not be stably maintained along boundaries with high curvature \cite{guillamat_2022_IntegerTopologicalDefects}. Second, considering that larger culture domains require more time to reach confluence and develop nematic order, geometries with smaller inner obstacles such as Patterns 1 and 2 may not have provided sufficient time for the annihilation of defect pairs, as evidenced by the presence of more defects in Patterns 1 and 2 than in  Patterns 3 and 4 (Fig.~\ref{fig:fig2}). Because the nematic correlation length $\xi_{nn}$ has been reported to reach $\xi_{nn}\approx 190$~\textmu m approximately 30 h after cell seeding in the case of unconfined C2C12 myoblast monolayers~\cite{guillamat_2022_IntegerTopologicalDefects}, the sizes of Patterns 1 and 2 were likely too large relative to this correlation length to develop high nematic order in confined cells, thereby suppressing defect pair annihilation. These considerations suggest that the proposed theory should accommodate alternative boundary conditions and domain sizes, which will be investigated in our future studies.

Our experimental results for Pattern 3 also showed multiple stable defect configurations (Patterns 3A and 3B in Fig.~\ref{fig:fig3b}), which corresponded to Cases B, C, and D in the numerical analysis. According to the elastic energy analysis (Fig.~\ref{fig:fig9}), the energy differences among the four cases were small; however, we did not observe any defects near the horizontal axis in any of the 97 images for Pattern 3, meaning that Case A was not observed experimentally. One possible reason for the lack of defects near the horizontal axis for Pattern 3 stems from the early-stage organization of the director field. In the time-lapse observations, cells between the two inner obstacles rapidly aligned along the vertical direction at an early stage, driven by tangential anchoring on the inner-obstacle surfaces (Fig.~\ref{fig:fig1}b). Defect pairs then formed in the open areas near the vertical axis. Such temporal dynamics of cell alignment imply that defects were rarely formed near the horizontal axis, because of the highly ordered nematic alignment established between the obstacles at an early stage. These results suggest that the stable defect configurations are governed not only by the energy landscape but also by the initial state of the director field. Therefore, our future work will focus on improving the proposed theoretical framework by incorporating initial defect distributions and the stochastic dynamics of defect configurations. 

\section{Conclusion}

We investigated, both experimantally and theoretically, the stable configurations of two $-1/2$ defects in cell monolayers confined to triply connected domains with an Euler characteristic of  $-1$.
The central finding of this work is that the size of the internal obstacles controls the symmetry of the stable configuration.
Specifically, the most stable configuration shifts continuously from a horizontal, through off-axis, to a vertical configuration as the obstacle size increases.
The defect positions observed experimentally agreed with these theoretical predictions to within 60~\textmu m, demonstrating that our theoretical framework enables accurate prediction of stable defect configurations and their stability prior to experiments.

Although existing studies on domains with non-negative Euler characteristics have shown that boundary curvature determines stable defect positions, the present study demonstrates that interactions between $-1/2$ defects and obstacles generate multiple critical points of the Frank elastic energy, despite the vertical symmetry of the domain considered here. In particular, diagonal defect configurations were frequently observed for the intermediate obstacle size, suggesting that cell contraction in such configurations may generate torsional force or torques. This observation leads to the hypothesis that cellular tissues control the nematic order of cell alignment via the size of internal holes to generate various complex force patterns during morphogenesis or organ movements. By analytically linking geometry-containing obstacles to defect configurations, this work provides a fundamental theoretical framework for exploring the geometrical and topological roles of the Euler characteristic in cell alignment and mechanics.

\section*{Acknowledgements}
The second author was supported by a JSPS Postdoctoral Fellowship (Grant Number JP24KJ0041) and JSPS KAKENHI (Grant Number JP26K17030). The third author was
supported in part by JSPS KAKENHI (Grant Number JP23H00086). The last author was
supported in part by JSPS KAKENHI (Grant Number JP23H00086 and JP26K00897) and  JST, PRESTO (Grant Number JPMJPR24KB).

\appendix

\section*{Appendix}
\label{app:methods}

\subsection{Fabrication of PDMS microwells}
PDMS microwells for cell cultures were fabricated by conventional soft lithography described in our previous study, with a slight modification \cite{miyazako_2024_PredictiveModelSpatialb}. First, SU-8 molds were fabricated as follows. SU-8 dry film resist of thickness 45~\textmu m (SU-8 3045CF DFR (WF1), Nippon Kayaku, Japan) was laminated onto a silicone substrate at $75\,^\circ\mathrm{C}$. The substrate was then exposed to UV-light patterns for 50~s by a maskless lithography system (PALET, NEOARK, Japan) equipped with a $2\times$ objective lens. After a 3-min bake at $55\,^\circ\mathrm{C}$ and 5-min bake at $95\,^\circ\mathrm{C}$, the substrate was developed and rinsed with SU-8 developer (Kayaku Advanced Materials, USA).

To fabricate PDMS microwells from the prepared SU-8 molds, we poured a mixture of a PDMS base elastomer and curing agent (SILPOT184, DuPont Toray Specialty Materials, Japan) with weight ratio of 10:1 onto SU-8 substrates in a plastic Petri dish. After a 3-h bake at $60\,^\circ\mathrm{C}$, the PDMS was removed from the mold and cut to fit into a 35-mm-diameter dish. The fabricated PDMS microwells were UV-sterilized for 1 h. 

To enhance the cell attachment of the fabricated PDMS substrates, the PDMS microwells were hydrophilized for 5 s by vacuum plasma (YHS-R, SAKIGAKE-Semiconductor, Japan) and coated with fibronectin solution (063-05591, Fujifilm Wako Pure Chemical Corporation, Japan) diluted 1:100 in D-PBS(--) (14249-95, NACALAI TESQUE, Japan), followed by incubation at $37\,^\circ\mathrm{C}$ for 1 h.

\subsection{Cell culture and observation}
Mouse myoblast C2C12 cells (provided by RIKEN BioResource Center, RCB0987) were cultured in Dulbecco's modified Eagle medium (low glucose, 08456-65, NACALAI TESQUE, Japan) with 10\% fetal bovine serum (S-FBS-NL-015, Serana Europe, Germany) and 1\% penicillin--streptomycin (26253-84, NACALAI TESQUE, Japan) at $37\,^\circ\mathrm{C}$ under 5\% CO$_2$.
Cells were passaged using trypsin-EDTA (35554-64, NACALAI TESQUE, Japan) for cell detachment and seeded onto the fibronectin-coated PDMS microwells. For observation of the cells, we used a benchtop imaging system equipped with a $10\times $ objective lens and an incubator for maintaining the temperature and moisture level. For time-lapse imaging, the dishes were placed in the imaging system 1 day after seeding. The phase-contrast microscope images were acquired every 30 min for 24 h.

\subsection{Image processing and statistical analysis}
\label{app:image}
Phase-contrast images were analyzed with FIJI/ImageJ~\cite{schindelin_2012_FijiOpensourcePlatform} and the OrientationJ plugin~\cite{puspoki_2016_TransformsOperatorsDirectional} to obtain the local alignment angle $\phi(x,y)$ and the scalar order parameter $S$ based on the procedure described in our previous study~\cite{miyazako_2024_PredictiveModelSpatialb}.
The window size for computing $S$ was set to \,$60\times60\,$ pixels ($\approx 28.8\times 28.8$~\textmu m) and local minima of $S$ were extracted as candidates for defects.
The topological charge $q$ was then determined by integrating the variation of the angle along a circle $\Gamma$ centered at each candidate point with a radius of $15$ pixels as follows:
\begin{equation}
q=\frac{1}{2\pi}\oint_{\Gamma}d\phi.
\end{equation}
Finally, candidates with $q\approx\pm\tfrac{1}{2}$ were identified as $\pm 1/2$ defects.

To determine the most probable defect positions in the experimental data  (Figs.~\ref{fig:fig3}a and \ref{fig:fig3b}a), we calculated the probability density using a two-dimensional Gaussian kernel density estimation with Scott's rule for bandwidth selection, as implemented in SciPy~\cite{virtanen_2020_SciPy10Fundamental}.
The density was evaluated on a $200 \times 200$ grid spanning the data domain, and the grid point with the maximum density was taken as the most probable defect position.

In Figs.~\ref{fig:fig3}c, e and \ref{fig:fig3b}c, the variances of the polar angle and the radial distance from the origin are compared between groups using Levene's test with the median, as implemented in SciPy~\cite{virtanen_2020_SciPy10Fundamental}.

\subsection{Numerical search for critical points of the Frank elastic energy}
\label{app:optimization}

The critical points of the Frank energy, where $\mathrm{f}_1 = \mathrm{f}_2 = 0$, under the symmetry constraints were numerically searched for as follows. For configurations with the defect pair on a symmetry axis (Cases A and B), $(z_1, z_2)$ reduces to a single variable; thus, the zeros of the relevant force components ($\mathrm{Re}\,\mathrm{f}_1$ or $\mathrm{Im}\,\mathrm{f}_1$) were computed by bisection at each $r$ with a step size of $0.01$.

For off-axis configurations (Cases C and D), $\mathrm{f}_1 = \mathrm{f}_2 = 0$ was solved with a Newton-type solver implemented in MATLAB\textregistered (\texttt{fsolve}). First, the solution for $r = 0.25$ was obtained by sampling $10^4$ random points of $z_1$ in the first quadrant and running Newton iterations from the $60$ points with the smallest $|\mathrm{f}_1|$. We then accepted solutions off both axes ($|\mathrm{Re}\,z_1|, |\mathrm{Im}\,z_1| > 10^{-6}$) with $|\mathrm{f}_1|, |\mathrm{f}_2| < 10^{-3}$. Using the accepted solution, we iteratively solved the solutions for other $r$ by changing $r$ with a step size of 0.01 and using the previous solution as the initial guess. 

To verify the stability of the obtained defect configurations under the symmetry assumptions, we classified each critical point as a minimum or a saddle point based on the signs of the real parts of the eigenvalues of the $4\times4$ Jacobian matrix, which was calculated numerically using central differences with a step size of $10^{-6}$. 

\bibliographystyle{apsrev4-2}
\bibliography{bibliography}

\end{document}